\documentclass[aps,prb,twocolumn,superscriptaddress]{revtex4-2}
\usepackage{makeidx}\makeindex
\usepackage{amsmath}
\usepackage{amssymb}
\usepackage{amsthm}
\usepackage{graphicx} %for pdfLatex
\usepackage{color}
\usepackage{bm}

\newcommand{\Tc}{T_\mathrm{c}}
\newcommand{\kB}{k_\mathrm{B}}

\newcommand{\degree}{$^\circ$}
\newcommand{\ohm}{$\mathrm{\Omega}$}

\newcommand{\roottimesroot}[2]{$\sqrt{#1}\times\sqrt{#2}$}

\begin{document}

%Title of paper
\title{Anisotropic transport of Josephson vortices in atomic-layer superconductors on vicinal surfaces}
\author{Wenxuan Qian}
%\email{QIAN.Wenxuan@nims.go.jp}
\affiliation{Research Center for Materials Nanoarchitectonics (MANA), National Institute for Materials Science, 1-1, Namiki, Tsukuba, Ibaraki 305-0044, Japan}
\affiliation{Graduate School of Science, Hokkaido University, Kita-10 Nishi-8, Kita-ku, Sapporo 060-0810, Japan}

\author{Yash Chauhan}
%\email{yash.chauhan@niser.ac.in}
\affiliation{School of Physical Sciences, National Institute of Science Education and Research, Jatni 752050, India}
\affiliation{Homi Bhabha National Institute, Training School Complex, Anushaktinagar, Mumbai 400094, India}

\author{Ryohei Nemoto}
%\email{nemoto@phys.sci.isct.ac.jp}
\affiliation{Research Center for Materials Nanoarchitectonics (MANA), National Institute for Materials Science, 1-1, Namiki, Tsukuba, Ibaraki 305-0044, Japan}
\affiliation{Department of Physics, School of Science, Institute of Science Tokyo, 2-12-1, Ookayama, Meguro-ku, Tokyo 152-8551, Japan}

\author{Keisuke Sagisaka}
%\email{SAGISAKA.Keisuke@nims.go.jp}
%\affiliation{Research Center for Materials Nanoarchitectonics (MANA), National Institute for Materials Science, 1-2-1, Sengen, Tsukuba. Ibaraki 305-0047, Japan}
\affiliation{Research Center for Materials Nanoarchitectonics (MANA), National Institute for Materials Science, 1-1, Namiki, Tsukuba, Ibaraki 305-0044, Japan}

\author{Shunsuke Yoshizawa}
%\email{YOSHIZAWA.Shunsuke@nims.go.jp}
%\affiliation{Research Center for Materials Nanoarchitectonics (MANA), National Institute for Materials Science, 1-2-1, Sengen, Tsukuba. Ibaraki 305-0047, Japan}
\affiliation{Research Center for Materials Nanoarchitectonics (MANA), National Institute for Materials Science, 1-1, Namiki, Tsukuba, Ibaraki 305-0044, Japan}

\author{Takashi Uchihashi}
\email[Corresponding author:]{UCHIHASHI.Takashi@nims.go.jp}
\affiliation{Research Center for Materials Nanoarchitectonics (MANA), National Institute for Materials Science, 1-1, Namiki, Tsukuba, Ibaraki 305-0044, Japan}
\affiliation{Graduate School of Science, Hokkaido University, Kita-10 Nishi-8, Kita-ku, Sapporo 060-0810, Japan}

\date{\today}

\begin{abstract}
Atomic steps have strong influences on surface two-dimensional superconductors. 
Josephson vortices formed at the atomic steps under magnetic fields may dominate transport phenomena at low temperatures, but its experimental verification is still lacking.
Here, we report the vortex transport properties of atomic-layer superconductor Si(111)-$(\sqrt{7}\times\sqrt{3})$-In with vicinal surfaces, for which Josephson vortices are directly observed by scanning tunneling microscopy.
A sharp drop in resistance with decreasing temperature $T$, detected under out-of-plane magnetic field $B$, reveals a distinctive anisotropy with respect to the atomic step direction.
The anisotropy of sheet resistance, proportional to that of vortex mobility, amounts to the order of $10^3$ at intermediate magnetic fields.
In the high-$T$ and low-$B$ region, Josephson vortices exhibit thermally excited creep motions with anisotropic activation energy $U_\mathrm{act}$.
A further increase in $B$ suppresses $U_\mathrm{act}$ toward zero anisotropically, resulting in one-dimensional pinning-free vortex flow at $0.10 \lesssim B \lesssim 0.20$ T.
At the lowest temperatures, the vortex motion is governed by quantum tunneling.
A $B$-$T$ phase diagram constructed based on these measurements reveals multiple regions characterized by directionally dependent vortex-transport mechanisms.
\end{abstract}

% insert suggested keywords - APS authors don't need to do this
%\keywords{}

%\maketitle must follow title, authors, abstract, and keywords

\maketitle
\section{Introduction}

Atomic-layer superconductors, realized using diverse surfaces, interfaces, and two-dimensional (2D) materials, have attracted extensive attention due to their distinctive and anomalous properties
\cite{Qin_Pb2ML,Zhang_PbIn1ML,Uchihashi_InR7R3Super,Ge_FeSeHighTc,Brun_PbSiDisorder,Lu_MoS2IsingSuper,Saito_MoS2IsingSuper,Tsen_BoseMetal,Ichinokura_SuperGraphene,Bollinger_SILaSrCuO,Saito_EDLSuper2D,Cao_TBGSuper,Ming_ChiralSuperSnSi,Uchihashi_2DSuperReview}. 
For example, they undergo superconductor-insulator transitions driven by disorder, magnetic field, and carrier density, thus providing an ideal platform for studying quantum phase transitions in 2D systems \cite{Haviland_BiSI,Saito_EDLSuper2D,Bollinger_SILaSrCuO,Cao_TBGSuper,Sato_StableVortices}.
Another remarkable phenomenon is an anomalously high in-plane critical magnetic field due to the breaking of the inversion symmetry and Rashba/Ising spin-orbit coupling \cite{Lu_MoS2IsingSuper,Saito_MoS2IsingSuper,Yoshizawa_DynamicRashba}.

Important geometric features that may govern the physics of atomic-layer superconductors are atomic steps that weakly couple the neighboring surface terraces.
They work as Josephson junctions and allow supercurrents to run over a macroscopic distance \cite{Uchihashi_InR7R3Super}. 
Under a magnetic field, such atomic steps accommodate Josephson vortices \cite{Yoshizawa_InVortex,Brun_PbSiDisorder,Roditechev_JV,Sato_SqueezedVortex}.
They are anomalous in that the vortex core is elongated along the atomic step and the superconducting order parameter within the core is largely sustained \cite{Yoshizawa_InVortex,Kawakami_JV,Blatter_VortexReview}.
These features are in striking contrast to those of Pearl vortices formed on 2D flat surfaces\cite{PearlVortex}.
Such vortices can be regarded as the 2D analogue of Josephson vortices induced within the layers of cuprate superconductors under an in-plane magnetic field \cite{Lee_BSCCOJVFlow,Ooi_JVFlow,Bae_JVTHzEmission,Koshelev_JVReview,Moler_JVImaging}

If the atomic steps are perfectly straight, Josephson vortices should flow freely along them because of translational symmetry, while their motion may be hindered by a pinning force in the perpendicular direction. 
Such an anisotropic transport of vortices can potentially affect the quantum phase transitions in 2D superconductors through the reduction of the effective dimensionality. 
They can also govern the thermal properties of the system, because an excess entropy is carried by a vortex core \cite{Solomon_NernstEffect,Vidal_EntropyFlux,Wang_HighTcNernstEffect,Tinkham_Textbook}.
Nevertheless, such a directional pinning effect of the atomic step has never been probed, and even its existence is not obvious because of a huge disparity between the typical length scales (atomic steps: $< 1$ nm, vortex core: $\sim 100$ nm).
Furthermore, the actual behavior of Josephson vortices can be sensitive to the morphological details of atomic steps, which generally break the translational symmetry. 
This calls for an experimental investigation of the transport properties of Josephson vortices.

Here, we report on highly anisotropic transport of Josephson vortices in an atomic-layer superconductor Si(111)-(\roottimesroot{7}{3})-In (referred to as (\roottimesroot{7}{3})-In below) \cite{Zhang_PbIn1ML,Uchihashi_InR7R3Super,Park_InSiDL,Shirasawa_SXRDInSi}, which is grown on a vicinal surface. 
The morphology of parallel atomic steps and the presence of Josephson vortices are confirmed using a scanning tunneling microscope (STM).
Transport properties of Josephson vortices in the directions parallel and perpendicular to the atomic steps are probed through four-terminal resistance measurements in a wide range of temperature $T$ and magnetic field $B$.
%The sheet resistances of the samples acquired at different magnetic fields $B$ reveal superconducting transitions at distinct temperatures $T$ with respect to the atomic step direction. 
The superconducting transition is detected as a sharp decrease in resistance under magnetic field $B$, revealing a distinctive anisotropy with respect to the atomic step direction.
The in-plane anisotropy of sheet resistance, proportional to vortex mobility, amounts to the order of $10^3$ at intermediate magnetic fields.
At relatively high $T$ and low $B$, the vortices show thermally excited creep motions with a distinctive in-plane anisotropy of activation energy $U_\mathrm{act}$.
$U_\mathrm{act}$ vanishes for the parallel direction at $B\approx 0.10$ T while remaining finite for the perpendicular direction up to $B\approx 0.20$ T, resulting in a one-dimensional (1D) pinning-free vortex flow in the intermediate fields.
At lower $T$, the vortex motions are strongly suppressed and governed by quantum tunneling.
A $B$-$T$ phase diagram constructed from these results clarifies the coexistence of directionally dependent vortex-transport mechanisms in multiple regions.

\section{Experimental}
\subsection{Instrumentation}
All experiments were conducted in ultrahigh vacuum (UHV) at a base pressure of $P< 1 \times 10^{-8}$ Pa.
STM measurements were carried out at 0.4 K. 
To secure bulk carriers and a tunneling current at this temperature, highly doped n-type Si substrates with resistivities $\rho <0.01 \ \mathrm{\Omega cm}$ were used.
Transport measurements were carried out separately in a home-built UHV instrument between 0.4 and 4.2 K.
For this purpose, non-doped Si substrates with resistivities $\rho >1000 \ \mathrm{\Omega cm}$ were used to quench the bulk carriers.
Other parameters were identical to those adopted for STM measurements. 
The effect of the substrate dopants on the superconducting properties of (\roottimesroot{7}{3})-In is negligible as revealed by the almost identical values of $\Tc$ for highly-doped and non-doped Si substrates  \cite{Zhang_PbIn1ML,Yoshizawa_DynamicRashba}).
To facilitate quantitative four-terminal resistance measurements, a current path was defined using a shadow mask technique \cite{Uchihashi_InR7R3Super,Yoshizawa_DynamicRashba}.
For details, see Methods in Supplemental Material \cite{SM_JVtransport}.

\subsection{Sample Design}

Figure~1(a) shows the side view of the schematic sample geometry in the present work.
The Si substrate has a vicinal (111) surface with the normal direction tilted by 0.2 \degree\ toward the $[\bar{1} \bar{1} 2]$ orientation.
Assuming that the atomic steps are mostly of the monatomic height ($h_0 = 0.31$ nm), the surface is composed of parallel terraces running in the $[1 \bar{1} 0]$ direction with an average width of 89 nm.
Such a regular atomic step array can be realized by adopting a special protocol of substrate cleaning  \cite{RegularSteps_APL,SM_JVtransport}.
The Si(111) surface is covered by In atomic bilayers to form (\roottimesroot{7}{3})-In, the structure model of which has been established previously \cite{Park_InSiDL,Shirasawa_SXRDInSi}.
It is a representative atomic-layer superconductor with a transition temperature of $\Tc \approx 3.1$ K \cite{Zhang_PbIn1ML,Uchihashi_InR7R3Super,Yoshizawa_DynamicRashba}.
The emergence of Josephson vortices at atomic steps on this surface has been established in previous work \cite{Yoshizawa_InVortex}.

%=================================================
% Figure 1
%=================================================
\begin{figure}
\includegraphics[width=7.5cm]{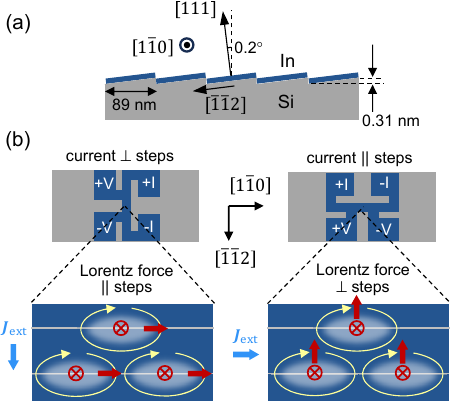}
\caption{Schematic drawing of the sample design. (a) Side view of the sample geometry. The Si(111) substrate has a vicinal surface with the normal direction tilted by 0.2 \degree\ toward $[\bar{1} \bar{1} 2]$ orientation. The surface is composed of parallel terraces running in the $[1 \bar{1} 0]$ direction, separated mostly by monatomic steps. The Si(111) surface is covered by In atomic bilayers to form (\roottimesroot{7}{3})-In. (b) Top view of the sample geometry adopted for transport measurements.
The upper panels illustrate linear current paths in two configurations, which are connected to voltage/current electrodes for four-terminal resistance measurements. The lower panels are microscopic views within the current paths. The cores of Josephson vortices are represented by white-shaded ovals, which are accompanied by circulating supercurrents (yellow lines) and quantum magnetic fluxes (red crossed circles). The red arrows show the directions of the Lorentz force exerted by external currents $J_\mathrm{ext}$. In the left panels, the Lorentz force is applied parallel to the atomic steps (parallel configuration), while perpendicular in the right panels (perpendicular configuration).
}
\label{Fig1}
\end{figure}

Figure~1(b) shows the top view of the sample design adopted for transport measurements.
The upper panels illustrate current paths with a size of $0.3 \times 1.2 \ \mathrm{mm^2}$ in two configurations, which are connected to voltage/current electrodes for four-terminal resistance measurements.
The lower panels show microscopic views within the current paths, where Josephson vortices  are created along the steps under an out-of-plane magnetic field.
The core size of the Josephson vortex in (\roottimesroot{7}{3})-In perpendicular to the steps is about $80$--$100$ nm \cite{Yoshizawa_InVortex}, which is approximately equal to the average terrace width of 89 nm.
This scale matching is intended to enhance the pinning effect on Josephson vortices and simultaneously to reduce the area of flat terraces available for Pearl vortices. 
In the left panels, a current runs in the direction perpendicular to the atomic steps.
Since the Lorentz force is exerted on vortices along the steps, vortices are expected to move easily in this direction.
This vortex motion will be detected by a large sheet resistance $R_\mathrm{sheet}$ of the sample through a relation \cite{Tinkham_Textbook}
\begin{equation}
%R_\mathrm{sheet} = B\Phi_0/\eta \propto \mu_{\phi},
R_\mathrm{sheet} = B\Phi_0 \mu_{\phi},
\label{eq:R-mobility_relation}
\end{equation}
%where $\Phi_0$ is the magnetic flux quantum, $\eta$ is the viscosity coefficient for vortices, and $\mu_{\phi}$ is the mobility of vortices.
where $\Phi_0(= h/2e)$ is the magnetic flux quantum and $\mu_{\phi}$ is the mobility of vortices.
$\mu_{\phi}$ is defined here by $\mu_{\phi}=-\langle v_{\phi} \rangle/f_\mathrm{L}$, where $\langle v_{\phi} \rangle$ is the average velocity of a vortex and $f_\mathrm{L}$ is the Lorentz force.
We note that Eq.~(\ref{eq:R-mobility_relation}) is conventionally expressed using viscosity $\eta (=\mu_{\phi}^{-1})$ for a continuous vortex flow. 
In the right panel of Fig.~1(b), by contrast, a current runs in the direction parallel to the steps.
In this case, vortices feel the Lorentz force in the direction perpendicular to the atomic steps, and thus are impeded by them.
This should result in a low vortex mobility $\mu_{\phi}$, resulting in a low resistance $R_\mathrm{sheet}$ according to Eq.~(\ref{eq:R-mobility_relation}).
In the following, the former and latter setups are called parallel and perpendicular configurations, respectively; \textit{i.e.,} the corresponding directions refer to those of vortex motions.

\section{Results and Discussion}
\subsection{STM Measurements}

%=================================================
% Figure 2
%=================================================
\begin{figure}
\includegraphics[width=8.5cm]{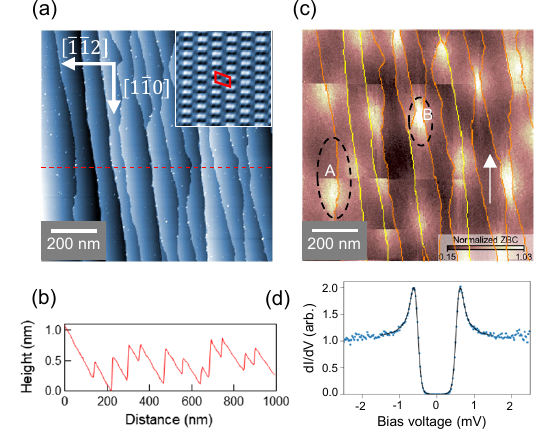}
\caption{STM characterization of the sample surface. (a) STM topograph image with a size of $1000 \times 1000 \ \mathrm{nm^2}$ with a sample bias voltage $V=100$ meV and a tunneling current $I=5$ pA. The arrows indicate the crystalline orientations of the surface. The inset is a magnified image with a size of $10 \times 10 \ \mathrm{nm^2}$ ($V = 500$ mV and $I = 50$ pA). The red parallelogram indicates the unit cell of (\roottimesroot{7}{3})-In. (b) Height profile taken along the dashed line in Fig. 2(a). (c) ZBC image acquired over the same area of Fig.~2(a) (Setpoint: $V=20$ meV and $I = 500$ pA). The orange-colored and yellow lines indicate the locations of monatomic and diatomic steps, respectively. A brighter color corresponds to a higher intensity of ZBC signal. The vertical arrow indicates the sub-scanning direction of STM imaging.
(d) $dI/dV$ spectrum acquired on a terrace of (\roottimesroot{7}{3})-In (blue dots, Setpoint: V = 10 mV and I = 500 pA), signifying the appearance of a superconducting energy gap. The solid line is a fit to the data using the Dynes formula.
All measurements were conducted at $T=0.4$ K. Magnetic field was set at $B=0.04$ T (a,c) and $B=0.00$ T (d).
}
\label{Fig2}
\end{figure}

Before carrying out transport measurements, we first characterized the surface morphology of a (\roottimesroot{7}{3})-In sample using an STM. 
As mentioned above, the sample was prepared using a highly doped Si substrate with the same protocol as adopted for transport measurements.
Figure~2(a) shows a representative STM topograph with an area of $1000 \times 1000 \ \mathrm{nm^2}$. 
The image reveals an array of atomic steps running approximately in the $[1 \bar{1} 0]$ direction.
A line profile taken along the red dashed line shows 12 terraces with an average width of about 80 nm, accompanied by seven monatomic steps (height: $h_0 = 0.31$ nm) and four diatomic steps (height: $2h_0 = 0.62$ nm) (Fig.~2(b)).
%The inconsistency of the apparent step heights in the line profile is due to a large effective radius of the tip.
These observations confirm that the surface morphology of the actual sample approximately follows the design in Fig.~1(a). 
The inset of Fig.~2(a) shows a magnified image with an area of $10 \times 10 \ \mathrm{nm^2}$, revealing a perfectly ordered In atomic layers \cite{Uchihashi_InR7R3Super,Kraft_InSurfaces}.
The red parallelogram indicates the unit cell of (\roottimesroot{7}{3})-In.

Superconductivity and vortex formation in (\roottimesroot{7}{3})-In were also confirmed through STM measurements.
Figure~2(d) shows a $dI/dV$ spectrum acquired on a surface terrace at $T=0.4$ K and $B=0$ T (blue dots), signifying the appearance of a superconducting energy gap.
Our analysis using the Dynes formula gives a nearly perfect fit (black line), giving an energy gap of $\Delta =0.552$ meV
%, together with the effective temperature of the sample $T_\mathrm{eff} = 0.555$ K, energy broadening due to quasiparticle lifetime $\Gamma =0.001$ meV 
%(black solid line) 
\cite{Dynes_QPLifetime}.
Assuming $\Tc=3.1$ K, the ratio of $\Delta/\kB\Tc=2.06$ is almost equal to a previously reported value of 2.08 \cite{Zhang_PbIn1ML}, while it is slightly larger than $\Delta/\kB\Tc =1.76$ predicted by the BCS theory. 
The zero-bias conductance (ZBC) mapping taken at $B = 0$ T shows that ZBC is uniformly suppressed nearly to zero including at the step edges, indicating homogeneous development of superconducting energy gap (see Fig. S1 of Supplemental Materials \cite{SM_JVtransport}). 
The absence of significant disorder helps establish superconductivity at least down to $T = 0.4$ K, which will also be shown by transport measurements below.
Figure~2(c) displays a ZBC image acquired at $T=0.4$ K and $B=0.04$ T over the same area of Fig.~2(a).
The orange-colored and yellow lines indicate the locations of monatomic and diatomic steps, respectively. 
A brighter color in the image corresponds to a higher intensity of ZBC signal and thus to a higher local density of states at the Fermi level.
The elongated bright regions straddling atomic steps are assigned to the cores of Josephson vortices.
Vortices prefer to take these positions because the recovery of superconducting order parameter at a Josephson vortex core leads to an increase in superconducting condensation energy gain \cite{Yoshizawa_InVortex,Kawakami_JV,Blatter_VortexReview}. 

Interestingly, most of the Josephson vortices in Fig.~2(c) appear cut in the middle of scanning (\textit{e.g.}, see Feature \textit{A}). 
This means that Josephson vortices emerge and vanish in the scanning area abruptly, suggesting a high vortex mobility along the steps as expected.
Here, the STM image was acquired by scanning horizontally, while the scanning line was translated vertically (indicated by the arrow).
The fact that several vortices on the same line moved simultaneously indicates a collective motion of vortices triggered by a single event. 
The mechanism of the vortex motion cannot be determined from the STM measurements; it could be driven by quantum tunneling at the lowest temperatures, as indicated by the transport measurements (see Sec. III-C), or by a vortex-STM tip interaction exerted during scanning.

We note that Fig.~2(c) includes bright-colored regions located within terrace constrictions, which are not disturbed by scanning (\textit{e.g.} see Feature \textit{B}).
These are ascribed to vortex cores immobilized at these constrictions.
As such, they should not affect transport properties of the overall system, at least at low magnetic fields where vortices are spatially separated.
In the following discussions, we will focus on the effects of Josephson vortices.

\subsection{Magneto-transport Measurements}

%=================================================
% Figure 3
%=================================================
\begin{figure}
\includegraphics[width=8.5cm]{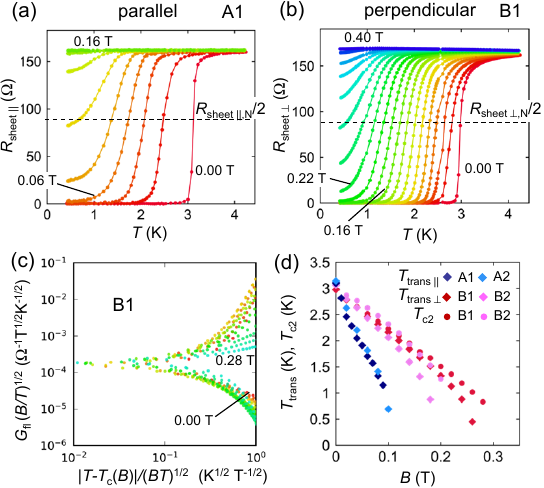}
\caption{Temperature-dependent sheet resistance $R_\mathrm{sheet}$, which is directly proportional to the mobility of vortices, measured at different magnetic fields. (a) $R_\mathrm{sheet\parallel}$ (Sample A1) for magnetic fields from $B=0.00$ to 0.16 T with an increment of 0.02 T. The dashed line indicates the $R_\mathrm{sheet\parallel}=R_\mathrm{sheet\parallel,N}/2$, from which $T_\mathrm{trans\parallel}$ is determined. (b) Analogous plot of $R_\mathrm{sheet\perp}$ (Sample B1) for magnetic fields from $B=0.00$ to 0.40 T with an increment of 0.02 T. (c) Ullah-Dorsey scaling analysis applied to the $R_\mathrm{sheet\perp}$ (Sample B1) from $B=0.00$ to 0.28 T with an increment of 0.02 T. $T_\mathrm{c2}$ is determined from the analysis. (d) $T_\mathrm{trans\parallel}$, $T_\mathrm{trans\perp}$, and $T_\mathrm{c2}$ plotted as a function of $B$ for all samples.
}
\label{Fig3}
\end{figure}

We now move on to the results of transport measurements under out-of-plane magnetic fields.
Four samples were prepared in total; Samples A1 and A2 for the parallel configuration (left panels of Fig.~1(b)) and Samples B1 and B2 for the perpendicular configuration (right panels of Fig.~1(b)).
The sheet resistances for the former and the latter are denoted as $R_\mathrm{sheet\parallel}$ and $R_\mathrm{sheet\perp}$, respectively.
The symbols for the parallel ($\parallel$) and perpendicular ($\perp$) configurations will be omitted when the description applies to both of them. 
The normal-state values measured at $T=4.2$ K are denoted as $R_\mathrm{sheet,N}$.

Figure~3(a) shows the sheet resistance $R_\mathrm{sheet\parallel}$ of Sample A1 as a function of temperature $T$ at different magnetic fields from $B=0.00$ to 0.16 T with an increment of 0.02 T.
For $B=0.00$ T, $R_\mathrm{sheet\parallel}$ plummets toward zero around $T=3.0$ K, signifying a superconducting transition \cite{Uchihashi_InR7R3Super,Yoshizawa_DynamicRashba}.
A small but finite resistance of $\sim 0.1$ \ohm\  remains even at sufficiently low temperatures, probably due to a stray magnetic field within the cryostat.
It could also be due to vortices and antivortices created during the superconducting transition through the Kibble-Zurek mechanism  \cite{Campo_ReviewKibbleZurek}. The determination of the exact cause is, however, beyond the scope of the present study.

This transition is rapidly suppressed by applying magnetic field.
At the lowest temperature of 0.4 K, $R_\mathrm{sheet\parallel}$ starts to deviate from zero around $B=0.06$ T, finally reaching 160 \ohm\ ($\approx R_\mathrm{sheet\parallel,N}$) at $B=0.16$ T.
The same experiment was carried out for Sample B1 from $B=0.00$ to 0.40 T (Fig.~3(b)).
In this case, $R_\mathrm{sheet\perp}$ is less sensitive to magnetic field; at the lowest temperature of 0.4 K, $R_\mathrm{sheet\perp}$ stays at less than 0.1 \ohm\ up to $B=0.16$ T.
This clearly indicates that a large portion of the sample is still covered by superconducting regions while vortices created by magnetic field can hardly move in the perpendicular direction.
By contrast, under the same condition, the vortices are highly mobile in the parallel direction, as signified by $R_\mathrm{sheet\parallel} \approx R_\mathrm{sheet\parallel, N}$ of Sample A1.
According to Eq.~(\ref{eq:R-mobility_relation}), $R_\mathrm{sheet\parallel}/R_\mathrm{sheet\perp}>10^3$ at $T=0.4$ K and $B=0.16$ T means that $\mu_{\phi\parallel}/\mu_{\phi\perp}>10^3$, where $\mu_{\phi\parallel}$ and $\mu_{\phi\perp}$ are vortex mobilities in the parallel and perpendicular directions, respectively.
$R_\mathrm{sheet\perp}$ exhibits a marked increase only above $B= 0.22$ T, reaching 170 \ohm\ ($\approx R_\mathrm{sheet\perp,N}$) at $B=0.40$ T.
The same experiments were conducted using Samples A2 and B2, confirming the reproducibility of the general behaviors (see Fig. S2 of Supplemental Material \cite{SM_JVtransport}).

Here, we define the transition temperature $T_\mathrm{trans\parallel}$ for the parallel configuration at a magnetic field $B$ from a relation $R_\mathrm{sheet\parallel}(T_\mathrm{trans\parallel})= R_\mathrm{sheet\parallel, N}/2$ (see the dashed line in Fig.~3(a)).
We note that $T_\mathrm{trans\parallel}$ for $B\neq 0$ reflects temperature-induced changes in vortex mobility $\mu_{\phi\parallel}$ rather than a thermodynamic superconducting transition.
Likewise, $T_\mathrm{trans\perp}$ is defined for the perpendicular configuration from $R_\mathrm{sheet\perp}(T_\mathrm{trans\perp})= R_\mathrm{sheet\perp, N}/2$ (see the dashed lines in Fig.~3(b)).
In this case, $T_\mathrm{trans\perp}$ corresponds to a thermodynamic transition, because superconductivity is broken at the mean-field level near $T_\mathrm{trans\perp}$.

The mean-field transition temperature under a magnetic field $B$ can be determined more systematically using a scaling analysis based on the Ullah-Dorsey theory, which is denoted as $T_\mathrm{c2}(B)$ below \cite{UllahDorsey_Scaling,Saito_Griffiths,Sato_StableVortices}.
According to the theory, $R_\mathrm{sheet\perp}$ measured as a function of $T$ and $B$ satisfies the following scaling relation:
\begin{equation}
G_\mathrm{fl} (B/T)^{1/2} = f\left( \frac{T-T_\mathrm{c2}(B)}{(BT)^{1/2}} \right),
\label{eq:Ullah-Dorsey}
\end{equation}
where $G_\mathrm{fl}\equiv R_\mathrm{sheet\perp}^{-1}-R_\mathrm{sheet\perp, N}^{-1}$ is a conductance due to superconducting fluctuations near $T_\mathrm{c2}$, and $f$ is a scaling function.
A series of $T_\mathrm{c2}(B)$ was determined by manually adjusting them in such a way that the curves of $G_\mathrm{fl} (B/T)^{1/2}$ plotted as a function of  $(T-T_\mathrm{c2}(B))/(BT)^{1/2}$ showed the best collapse for the whole data set.
Figure~3(c) shows the result of such an analysis, showing that all data roughly collapse on a single curve. 
The deviation from the ideal scaling behavior may be attributed to the in-plane anisotropy of the system, while the theory assumes an isotropic 2D superconductor. 
Nevertheless, the convergence of the curves is excellent near $T_\mathrm{c2}$, probably because significant overlap of vortices in both directions makes the system less anisotropic near $T_\mathrm{c2}$.
%We note that this scaling analysis is not applicable to the transition in the parallel configuration, which is driven by the change in vortex mobility.

Figure~3(d) plots the results on $T_\mathrm{trans\parallel}$, $T_\mathrm{trans\perp}$, and $T_\mathrm{c2}$ as a function of $B$ for all samples.
As expected, $T_\mathrm{trans\perp}$ and $T_\mathrm{c2}$ show similar dependencies.
%The differences between them can be reduced by defining $T_\mathrm{trans\perp}$ by, \textit{e.g.}, $R_\mathrm{sheet\perp}(T_\mathrm{trans\perp})= 0.8R_\mathrm{sheet\perp, N}$, and thus have no significant meaning.
In the following, we use $T_\mathrm{c2}$ to represent the thermodynamic phase transition because it is uniquely determined from Eq.~(\ref{eq:Ullah-Dorsey}), while $T_\mathrm{trans\parallel}$ is subject to the numerical factor in its definition.
We find that $T_\mathrm{trans\parallel}$ is significantly lower than $T_\mathrm{c2}$.
This means that Josephson vortices are highly mobile only in the parallel direction within a wide range of $T_\mathrm{trans\parallel}<T<T_\mathrm{c2}$.

\subsection{Mechanisms of Josephson Vortex Transport}

%=================================================
% Figure 4
%=================================================
\begin{figure}[!t]
\includegraphics[width=8.5cm]{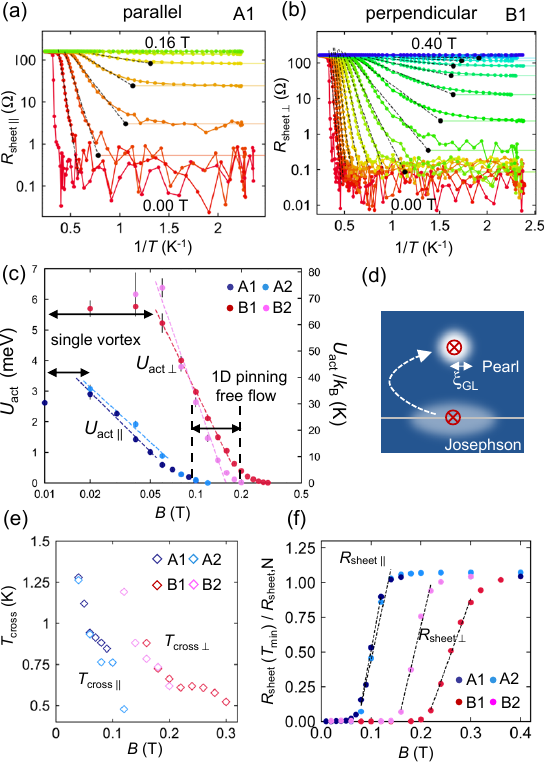}
\caption{Analysis on the Josephson vortex transport. (a) Arrhenius plots of the same $R_\mathrm{sheet\parallel}$ data of Fig.~3(a) (Sample A1). (b) Arrhenius plots of the $R_\mathrm{sheet\perp}$ data of Fig.~3(b) (Sample B1). In (a) and (b), the approximately straight lines in the high temperature regions indicate thermally activated transport of vortices, while the saturations at the lowest temperatures suggest quantum tunneling of vortices. The black dots correspond to the crossover temperatures $T_\mathrm{cross\parallel}$ and $T_\mathrm{cross\perp}$. (c) Activation energies $U_\mathrm{act\parallel}$ (Samples A1 and B1) and $U_\mathrm{act\perp}$ (Samples A2 and B2) plotted as a function of $B$. The horizontal axis is set in the log scale. As a reference, the scale for $U_\mathrm{act}/\kB$ is set on the right axis. The $B$ regions corresponding to the single-vortex excitation and 1D pinning-free vortex flow are indicated by the arrows. (d) Schematic picture of the excitation of a Josephson vortex into a Pearl vortex. The white areas represent the vortex cores. (e) $T_\mathrm{cross\parallel}$ and $T_\mathrm{cross\perp}$ plotted as a function of $B$. (f) $R_\mathrm{sheet}(T_\mathrm{min})/R_\mathrm{sheet,N}$ plotted as a function of $B$.
The linear increase above the threshold and the saturation around $R_\mathrm{sheet,N}$ indicate pinning-free vortex flow.
}
\label{Fig4}
\end{figure}

The mechanisms of Josephson vortex transport are clarified as follows.
Figures~4(a) and 4(b) are Arrhenius plots of the same $R_\mathrm{sheet}$ data of Fig.~3(a) and 3(b); $R_\mathrm{sheet\parallel}$ (Sample A1) and $R_\mathrm{sheet\perp}$ (Sample B1) are replotted as a function of $1/T$, with the vertical axes set in the log scale.
The data points follow approximately straight lines in the high temperature regions.
This indicates thermally activated transport of vortices described by
\begin{equation}
R_\mathrm{sheet} \propto \exp{\left(-\frac{U_\mathrm{act}}{\kB T} \right)},
\label{eq:ThermalAtivation1} 
\end{equation}
where $U_\mathrm{act}$ is an activation energy for vortex motion.
The black dashed lines in Figs.~4(a) and 4(b) represent the fit to Eq.~(\ref{eq:ThermalAtivation1}) in the high-$T$ regions.
The transport data of Samples A2 and B2 were analyzed in the same way (see Fig. S3 of Supplemental Material \cite{SM_JVtransport}).

Activation energies $U_\mathrm{act\parallel}$ and $U_\mathrm{act\perp}$ determined in this way are plotted in Fig.~4(c) for all samples, where the horizontal axis is set in the log scale.
As a reference, the scale for $U_\mathrm{act}/\kB$ is set on the right axis.
The data points taken from the same configuration (A1/A2: parallel, B1/B2: perpendicular) show a reasonable agreement, demonstrating the reproducibility of the present work. 
$U_\mathrm{act\parallel}$ (Samples A1 and A2) stays around 3 meV for $B\leq 0.02$ T, while $U_\mathrm{act\perp}$ (Samples B1 and B2) remains around 6 meV for $B\leq 0.05$ T, giving $U_\mathrm{act\perp}-U_\mathrm{act\parallel} \approx 2-4$ meV in this region.
The low density of vortices in this region limits the intervortex coupling, leading to nearly constant values of $U_\mathrm{act}$ due to single vortex excitations. 
The relation $U_\mathrm{act\parallel}<U_\mathrm{act\perp}$ reflects the fact that Josephson vortices can move more easily along the atomic steps.
Nevertheless, the motion requires a finite activation energy even in this direction, because of the non-ideal morphology of atomic steps (see Fig.~2(a)).

The difference between $U_\mathrm{act\parallel}$ and $U_\mathrm{act\perp}$ can be mostly attributed to a change in the core energy of a vortex.
Suppose that a Josephson vortex located at an atomic step is excited to move onto a terrace to form a Pearl vortex.
While the core of a Josephson vortex retains the superconducting condensation energy to a large extent, that of a Pearl vortex loses it because of strong suppression of the order parameter \cite{Yoshizawa_InVortex,Blatter_VortexReview,Tinkham_Textbook}.
This change in the energy can be roughly estimated as \cite{Tinkham_Textbook}
\begin{equation}
\Delta E_\mathrm{core}=\frac{1}{2}\rho(\epsilon_\mathrm{F})\Delta_0^2\pi \xi_\mathrm{GL}^2 d,
\label{eq:VortexCoreEnergy}
\end{equation}
where $\rho(\epsilon_\mathrm{F})$ is the density of states at the Fermi level, $\Delta_0$ is the superconducting energy gap, $\xi_\mathrm{GL}$ is the Ginzburg-Landau coherence length ($\approx$ the radius of the vortex core), and $d$ is the thickness of the bilayer of In(001).
Substituting $\rho(\epsilon_\mathrm{F})=2.0 \times 10^{28} \ \mathrm{eV^{-1}m^{-3}}$, $\Delta_0=1.76\kB\Tc =4.7\times 10^{-4} \ \mathrm{eV}$, $\xi_\mathrm{GL}=(\Phi_0/2\pi B_\mathrm{c2})^{1/2}=34$ nm, and $d=0.495$ nm into Eq.~(\ref{eq:VortexCoreEnergy}), one obtains $\Delta E_\mathrm{core}=4.0$ meV.
Here $\rho(\epsilon_\mathrm{F})$ and $d$ are taken from the bulk In values, and $B_\mathrm{c2}=0.28$ T is determined above for the lowest temperatures (see Fig.~3(d)).
Although Eq.~(\ref{eq:VortexCoreEnergy}) is not precise in terms of the numerical factor, it should be qualitatively valid. 
The result is in good agreement with $U_\mathrm{act\perp}-U_\mathrm{act\parallel} \approx 2-4$ meV for single vortex excitations.

A further increase in $B$ causes significant overlap between vortices, reducing pinning potential barriers for both directions.
In this case, the 2D collective pinning theory predicts the following equation: \cite{Feigelman_VortexPinning,Ephron_VortexFlow,Tsen_BoseMetal}
\begin{equation}
U_\mathrm{act}(B) = U_0 \ln{(B_0/B)},
\label{eq:CollectivePinning} 
\end{equation}
where $U_0$ is a pinning energy scale and $B_0$ is the field at which the activation energy vanishes and pinning-free vortex flow sets in.
The data of $U_\mathrm{act\parallel}$ for $B\geq 0.02$ T and $U_\mathrm{act\parallel}$ for $B \geq 0.05$ T can be well fitted by Eq.~(\ref{eq:CollectivePinning}), as shown by the straight dashed lines in Fig.~4(c).
The fitting analysis gives $B_0 = 0.08, 0.10$ T for samples A1 and A2, while $B_0 = 0.20, 0.16$ T for samples B1 and B2, respectively. 
Within the region of $0.1 \lesssim B \lesssim 0.2$, $U_\mathrm{act\parallel}$ vanishes, while $U_\mathrm{act\perp}$ remains finite, signifying a 1D pinning-free vortex flow along atomic steps.
We note that the derivation of Eq.~(\ref{eq:CollectivePinning}) assumes a collective pinning in a 2D system, which may not be rationalized in our anisotropic system. 
More appropriate forms of $B$-dependencies of $U_\mathrm{act\parallel}$ and $U_\mathrm{act\perp}$ are desirable for quantitative analysis.

The vortex transport is governed by a different mechanism at the lowest temperatures.
Figures~4(a) and 4(b) show that $R_\mathrm{sheet}$ tends to saturate for both directions.
The asymptotic values, denoted as $R_\mathrm{sheet}(T_\mathrm{min})$, were determined from the middle values of $R_\mathrm{sheet}$ acquired at the five lowest temperatures.
This $T$-independence strongly suggests that vortex motion between the pinning potential minima is governed by quantum tunneling.
It also points to the emergence of the anomalous metal phase, where liquid-like Cooper pairs and vortices, neither condensed nor localized, lead to an ohmic dissipation even in the limit of $T=0$ K \cite{Kapitulnik_AnomalMetal,Phillips_ReviewBoseMetal}.
Such an exotic phase has been reported for a variety of 2D superconductors, especially when the crystallinity is high and the normal sheet resistance is low \cite{Saito_EDLSuper2D,Tsen_BoseMetal,Ienaga_QVL}.
The present work suggests that a highly anisotropic anomalous metal can exist under the influence of atomic step arrays.
%Existence of anomalous metal in vicinal surface: not obvious because of and a direct S-I transition may occur.

The transition from thermal activation to quantum tunneling occurs around the crossover temperature $T_\mathrm{cross}$, which corresponds to the crossing point between the black dashed line and the horizontal line of $R_\mathrm{sheet}= R_\mathrm{sheet}(T_\mathrm{min})$ (see Figs.~4(a) and 4(b)).
In Fig.~4(e), $T_\mathrm{cross\parallel}$ and $T_\mathrm{cross\perp}$ are plotted as a function of $B$ for all samples, revealing a clear anisotropy.
Figure~4(f) displays $R_\mathrm{sheet}(T_\mathrm{min})$ normalized by $R_\mathrm{sheet,N}$ as a function of $B$.
They show a substantial increase above $B=0.1$ and $0.2$ T, respectively, around which $U_\mathrm{act} \approx 0$ meV sets in (see Fig. 4(c)).
The linear increase in $R_\mathrm{sheet}(T_\mathrm{min})$ above the threshold and the saturation around $R_\mathrm{sheet,N}$ indicates pinning-free vortex flow \cite{Tinkham_Textbook,Bardeen_VortexFlow,Ephron_VortexFlow,Saito_EDLSuper2D,Tsen_BoseMetal}.

\subsection{Phase Diagram}

%=================================================
% Figure 5
%=================================================
\begin{figure}
\includegraphics[width=8cm]{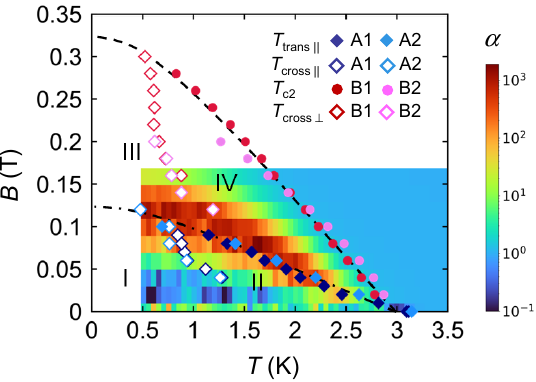}
\caption{$B$-$T$ phase diagram constructed from $T_\mathrm{trans\parallel}$ and $T_\mathrm{cross\parallel}$ of Samples A1/A2 as well as $T_\mathrm{c2}$ and $T_\mathrm{cross\perp}$ of Samples B1/B2. Regions I -- IV are characterized by different vortex-transport mechanisms : (I) anisotropic quantum creep, (II) parallel thermal creep and perpendicular quantum creep, (III) parallel pinning-free flow and perpendicular quantum creep, and (IV) parallel pinning-free flow and perpendicular thermal creep. The dashed and dash-dotted lines are eye guides for the expected boundaries. The anisotropy $\alpha$ defined by Eq.~(\ref{eq:Rsheet_anisotropy}) is also plotted in a color scale.
}
\label{Fig5}
\end{figure}

Figure~5 shows a $B$-$T$ phase diagram constructed from $T_\mathrm{trans\parallel}$ and $T_\mathrm{cross\parallel}$ of Samples A1/A2 as well as $T_\mathrm{c2}$ and $T_\mathrm{cross\perp}$ of Samples B1/B2. 
The line defined by a series of $T_\mathrm{c2}$ constitutes a thermodynamic phase boundary between the normal and superconducting states at the mean-field level, as mentioned earlier.
When the line is viewed as a variation of upper critical field $B_\mathrm{c2}$ as a function of $T$, $B_\mathrm{c2}$ first increases linearly with decreasing $T$ and tends to saturate at the lowest temperatures.
Theoretically, it is given by a relation $B_\mathrm{c2} =\Phi_0/2\pi \xi_\mathrm{GL}^2$.
%where $\xi_\mathrm{GL}$ is the GL coherence length.
The relations of $\xi_\mathrm{GL}\propto (1-T/\Tc)^{-1/2}$ near $\Tc$ and $\xi_\mathrm{GL} \to \mathrm{const.}$ as $T \to 0$ are consistent with our observation \cite{Tinkham_Textbook}.
The dashed line in Fig.~5 depicts the expected behavior of $B_\mathrm{c2}$ down to $T=0$. 
We note that the quantum Griffiths singularity was previously reported near $B_\mathrm{c2}$ at the lowest temperatures for highly crystalline 2D superconductors \cite{Xing_Griffiths,Saito_Griffiths}. 
It is interesting to clarify whether the same phenomenon is observed in the presence of atomic step arrays and Josephson vortices, which will be a forthcoming study.

The superconducting phase is primarily divided by the $T_\mathrm{trans\parallel}$ line into two regions.
The boundary given by $T_\mathrm{trans\parallel}$ is linear at high-$T$ region and tends to flatten as $T \to 0$ (dash-dotted line).
This is consistent with the fact that vortex motion is governed by quantum tunneling at the lowest temperatures and thus becomes insensitive to $T$.
In the low-$B$ region (denoted as I and II), vortex motion is limited by the pinning potential in both parallel and perpendicular directions.
%, leading to $R_\mathrm{sheet\parallel}\approx 0$ and $R_\mathrm{sheet\perp}\approx 0$.
By contrast, the high-$B$ region (denoted as III and IV) is featured with 1D vortex flow in the parallel direction.
% resulting in $R_\mathrm{sheet\parallel}\approx R_\mathrm{sheet\parallel,N}$ and $R_\mathrm{sheet\perp}\approx 0$.
%Thus, a dimensional crossover regarding the vortex transport behavior occurs around the $T_\mathrm{trans\parallel}$ line.
These regions are further divided by the $T_\mathrm{cross\parallel}$ and $T_\mathrm{cross\perp}$ lines.
In the high-$T$ side of the $T_\mathrm{cross\parallel}$ lines (Region II), the vortex transport in the parallel direction is governed by thermal activation, which is replaced by quantum tunneling in the low-$T$ side (Region I).
The same crossover occurs around $T_\mathrm{cross\perp}$ between Regions IV and III in the perpendicular direction.
Thus, Regions I, II, III, and IV in the phase diagram are characterized by the directionally dependent vortex-transport mechanisms.
%In Region I, the vortex transport is governed by anisotropic quantum creep motions, which are partially replaced by thermally activated creep in the parallel direction in Region II.
%In Region III, vortex flow is pinning-free in the parallel direction but is governed by quantum tunneling in the perpendicular direction.
%Finally, in Region IV, this quantum tunneling is replaced by thermal excitation while pinning-free vortex flow in the parallel direction remains.

The anisotropy of $R_\mathrm{sheet}$ is strongly enhanced at some parameter regimes as seen above.
To gain an overall view, we calculated anisotropy $\alpha$ from the experimental data using the following equation
\begin{equation}
\alpha =\frac{\sum_{\mathrm{A1, A2}} R_\mathrm{sheet\parallel}/R_\mathrm{sheet\parallel,N}}{\sum_{\mathrm{B1, B2}} R_\mathrm{sheet\perp}/R_\mathrm{sheet\perp,N}}.
\label{eq:Rsheet_anisotropy}
\end{equation}
Here, $R_\mathrm{sheet\parallel}$ and $R_\mathrm{sheet\perp}$ are normalized with respect to their normal state values, $R_\mathrm{sheet\parallel,N}$ and $R_\mathrm{sheet\perp,N}$, respectively, and are averaged over different samples. 
The calculation is limited to the region where the data of all samples are available.
$\alpha$ is plotted in a color scale within the $B$-$T$ phase diagram (Fig.~5).
The figure reveals that $\alpha$ is strongly enhanced along the $T_\mathrm{trans\parallel}$ line, amounting to the order of $10^3$.
%This is consistent with the fact that $R_\mathrm{sheet\parallel}\approx R_\mathrm{sheet\parallel,N}$ and $R_\mathrm{sheet\perp}\approx 0$ near this line.
Together with Eq.~(\ref{eq:R-mobility_relation}), the result indicates the anisotropy of vortex mobility $\mu_{\phi\parallel}/\mu_{\phi\perp}$ is also strongly enhanced to the same order.

\section{Conclusion}
We have investigated Josephson vortices in atomic-layer superconductors (\roottimesroot{7}{3})-In with vicinal surfaces, where the average separation of the atomic steps was of the same order of the vortex core size.
Our STM measurements not only directly visualized Josephson vortices located at atomic step, but also suggested that they were mobile along the steps as expected.
The transport properties of vortices in the directions parallel and perpendicular to the steps were clarified through four-terminal resistance measurements.
The $T$-dependence of sheet resistances $R_\mathrm{sheet}$ of the samples acquired at different magnetic fields $B$ revealed strong anisotropy of vortex transport properties; they manifested themselves as anisotropies in transition temperature $T_\mathrm{trans}$, activation energy $U_\mathrm{act}$, and crossover temperature $T_\mathrm{cross}$. 
Particularly, 1D pinning-free vortex flow was identified in the intermediate $B$ region.
The $B$-$T$ phase diagram constructed from these analyses, as well as mean-field transition temperature $T_\mathrm{c2}$, revealed multiple regions characterized with directionally dependent vortex-transport mechanisms.

As clarified in this work, the presence of atomic steps on a surface 2D superconductor leads to highly anisotropic transport of vortices, effectively reducing the dimensionality of the system.
This is likely to have strong influences on the quantum phase transitions in general, which are driven by disorder, magnetic field, and carrier density.
Particularly, the nature of the anomalous metal and the quantum Griffiths singularity may fundamentally be altered \cite{Saito_EDLSuper2D,Tsen_BoseMetal,Phillips_ReviewBoseMetal,Xing_Griffiths,Saito_Griffiths}.
It will be interesting to investigate whether the critical behavior of the phase transitions can be modified by the reduction of the effective dimensionality.
Another direction of future work will be to investigate anisotropic heat flow, which is expected because the vortex core accompanies an excess entropy.
Direct observation of anisotropic thermal transport may be possible through, for example, observation of the Nernst effect \cite{Solomon_NernstEffect,Vidal_EntropyFlux,Wang_HighTcNernstEffect,Ienaga_QVL}.
As demonstrated in this work, the directionality of heat flow may be significantly controlled by external parameters such as temperature and magnetic field, which may be used for future applications.
The present work lays a solid foundation for such investigations.

%(additional, discussions)
%vortex flow (flux flow) resistance: Bardeen-Stephan model, \cite{Tinkham_Textbook,Bardeen_VortexFlow}
%vortex pinning effect of atomic steps is not obvious; atomic-scale defects are generally very weak : Tinkham Sec. 5.4 \cite{Tinkham_Textbook} 
%2D collective pinning of vortex: Tinkham Chap.9
%transport of entropy : thermomagnetic effects (Ettingshausen Effect: transverse temperature gradient due to current) : Tinkham Sec. 5.5 \cite{Tinkham_Textbook}
%B-T phase diagram: similar to Helfand-Werthamer theory of Hc2 without spin effects (paramagnetic pair breaking and spin-orbit coupling). 
%Considering small magnetic field in the present experiment, the spin effect can be neglected. 

\nocite{Uchihashi_InR7R3Super,RegularSteps_APL}
%references in Supplemental Material

\vspace{10pt}

\textit{Acknowledgments}---This work was supported financially by JSPS KAKENHI (Grant Numbers 22H01961, 25K01670, 25H00867, 24K01351) and World Premier International Research Center (WPI) Initiative on Materials Nanoarchitectonics, MEXT, Japan.

\textit{Data availability}---The data that support the findings of this article are not publicly available. The data are available from the authors upon reasonable request.

%\begin{acknowledgments}

%\end{acknowledgments}

% Create the reference section using BibTeX:
%\bibliography{MyEndNoteLibrary}
\bibliography{MyEndNoteLibrary2}

\end{document}